\documentstyle [12pt,aps,amsfonts] {revtex}
\input epsf
\newcommand{\beq}{\begin{equation}}
\newcommand{\eeq}{\end{equation}}
\newcommand{\beqs}{\begin{eqnarray}}
\newcommand{\eeqs}{\end{eqnarray}}

\newcommand{\gsim}{\mathrel{\raisebox{-.6ex}{$\stackrel{\textstyle>}{\sim}$}}}

\begin{document}
\draft
 
\baselineskip 6.0mm
 
\tighten
 
\title{On ``Sub-Threshold'' Reactions Involving Nuclear Fission } 
 
\author{M. Goldhaber\thanks{email: goldhaber@bnl.gov}
\and R. Shrock\thanks{robert.shrock@sunysb.edu; 
on sabbatical leave from Yang Institute for
Theoretical Physics, State University of New York, Stony Brook through
July, 2000}}
 
\address{Physics Department \\
Brookhaven National Laboratory \\
P. O. Box 5000 \\
Upton, NY  11973-5000}
 
\maketitle
 
\vspace{10mm}
 
\begin{abstract}

We analyze reactions of several types that are naively below threshold but can
proceed because of the release of binding energy from nuclear fission and 
occasionally the formation of Coulombic bound states.  These reactions
include (i) photofission with pion production and (ii) charged current
neutrino-nucleus reactions that lead to fission and/or formation of a Coulomb
bound state of a $\mu^-$ with the nucleus of a fission fragment.  We comment on
the possible experimental observation of these reactions. 
 
\end{abstract}
 
\pacs{}
 
\vspace{16mm}
 
\newpage

\section{Introduction}

There are several types of reactions that can proceed even with
incident particles whose energies are apparently below threshold, because
of certain mechanisms that make the requisite energy available.  This
is true, in particular, for reactions in which the target is a
nucleus, denoted here by $(Z,A)$, where $Z$ denotes the number of
protons and $A$ denotes the mass number.  Here we explore reactions
involving heavy nuclei which can proceed, although the energy of the
incident particle is ``below threshold'', because the reaction
leads to the fission of the nucleus, thereby releasing a substantial
amount of energy, $E_{fiss.}  \sim 150$ MeV, which may be partially 
available to exceed a threshold.  We also consider the
small energy release due to the formation of a Coulomb bound state of
a $\mu^-$ with the nucleus or with a fission fragment. 

\section{Photofission with Pion Production}

Fission in heavy nuclei can be induced in several ways, in addition to
sometimes occurring spontaneously; these ways include bombardment by
slow (thermal and resonance) neutrons with energies from about 0.02 to
O(10) eV, by medium and fast neutrons with energies from keV to MeV,
by other hadronic projectiles such as protons, deuterons, and $\pi$
mesons, by photons, and by the capture of $\mu^-$'s (some reviews are
\cite{hyde,wilets}, a recent conference is \cite{fission98}, and a 
recent paper on photofission is \cite{tjl}). 
Typical fission energy barriers (measured, e.g., in fast neutron or
photofission reactions) are around 5 MeV.  In general, from the
roughly 0.1 eV widths of resonances observed in fission induced by
neutrons with energies between 0.2 and $10^2$ eV, it has been inferred
that the time taken for fission to occur is $t_{fiss.}  \sim 10^{-14}$
sec \cite{hyde}.

Fission induced by incident photons, photofission, has been extensively
studied.  The fissioning nucleus breaks into two fragments:
\beq
\gamma + (Z,A) \to (Z_1,A_1)+(Z_2,A_2)
\label{photofission}
\eeq
with $Z_1+Z_2=Z$, $A_1+A_2=A$. 
The fission daughter nuclei are typically produced in excited states and 
de-excite with prompt neutron and $\gamma$-ray emission.  
The cross section for photofission 
in a heavy element such as $^{238}$U rises from threshold for photon energies 
of about $E_\gamma \simeq 5$ MeV to a pronounced peak of about 100 mb at 
$E_\gamma \simeq 14$ MeV due to the excitation of
the giant dipole resonance \cite{gt}.  For higher photon energies, the cross 
section decreases, and then increases again for $E_\gamma \gsim 100 $ MeV.  
This second increase is interpreted as being due to the production of virtual
excited states of a nucleon in the nucleus, in particular, 
$\Delta(1232)$, which 
transfer energy and thereby catalyze the fission process \cite{hyde,js}.  

One may also consider photoproduction reactions yielding a pion in the final
state, including \beq \gamma + (Z,A) \to (Z,A)_{g.s.} + \pi^0
\label{gpi0el}
\eeq
\beq
\gamma + (Z,A) \to (Z \mp 1, A)_{g.s.} + \pi^\pm
\label{gpipel}
\eeq
where $g.s.$ means ground state, and the fission processes 
\beq
\gamma + (Z,A) \to (Z_1,A_1)+(Z_2,A_2) + \pi^0
\label{pi0}
\eeq
and
\beq
\gamma + (Z,A) \to (Z_1,A_1)+(Z_2,A_2) + \pi^\pm
\label{pipm}
\eeq
The $\pi^0$ decays to $\gamma\gamma$ with a mean 
lifetime of $\sim 0.8 \times 10^{-16}$ 
sec, so that the signature for reaction (\ref{pi0}) would be the detection of
the two fission fragments together with the two photons whose invariant 
mass reconstructs to that of the $\pi^0$.  The signature for reaction 
(\ref{pipm}) would be the detection of the two fission fragments together 
with the $\pi^\pm$, which, however, may undergo charge-exchange reactions in 
the nucleus, so that one would actually detect the diphoton signal from the 
decay of the $\pi^0$.  Before one takes account of the fission energy
the naive threshold ($thr.$) energy for the reaction (\ref{pi0}) is
\beq
E_{\gamma,0,thr.} = m_{\pi^0} + \frac{m_{\pi^0}^2}{2M(Z,A)} \ . 
\label{egammath0}
\eeq
With the definition of the difference in nuclear ground state (g.s.) 
energies
\beq
\Delta E_{Z \pm 1,Z;A} = E_{(Z \pm 1,A),g.s.}-E_{(Z,A),g.s.}
\label{deltae}
\eeq
the analogous naive
threshold photon energy for the reaction (\ref{pipm}) is 
\beq
E_{\gamma,\pm,thr.} = m_{\pi^\pm} + \frac{m_{\pi^\pm}^2}{2M(Z,A)} + 
\Delta E_{Z \pm 1,Z;A} \ . 
\label{egammathpm}
\eeq 
Of course, one must also take account of Fermi momentum and Pauli blocking
in the reaction kinematics.  Now consider a case where the available energy,
including the effect of the Fermi momentum of the struck nucleon, is below
threshold for pion production without fission.  However, with fission, pion
production might still occur because of the energy $E_{fiss.}$ may be partially
transferred by strong, as well as Coulombic, interactions, from the final state
fission fragments to the pion.

In the relevant range of photon energy, slightly below the naive threshold of
about 140 MeV for pion production, the total photofission cross sections on
$^{238}$U and $^{232}$Th are 
approximately 150 mb and 45 mb, respectively \cite{js,hyde}.  Since the
transfer of the fission energy to the pion is via the strong interaction, this
transfer should not significantly reduce the cross section.  However, for total
energies only slightly beyond the true threshold, there would be substantial
phase space suppression of the three-body reactions (\ref{pi0}) and
(\ref{pipm}).  Although the full reaction involves the integral over phase
space of the squared amplitude $|{\cal M}|^2$ 
for the reaction, one can obtain a rough
measure of the kinematic threshold suppression by considering the threshold 
dependence of the phase space factor by itself.  Recall that the usual
Lorentz-invariant $n$-body final state phase space integration is
\beq
R_n = (2\pi)^{4-3n}\int \Bigl [ \prod_{j=1}^n \frac{d^3p_j}{2E_j} \Bigr ] 
\delta(P-\sum_{j=1}^n p_j)
\label{rn}
\eeq
where $p_j^\lambda = (E_j, {\bf p}_j)$ denotes the 4-momenta of
the $j$'th final-state particle, and $P^\lambda=(\sqrt{s},0,0,0)$ in the 
center-of-mass-frame.  Let us denote the overall energy release as 
\beq
Q = \sqrt{s}-\sum_{j=1}^n m_j . 
\label{qval}
\eeq
Near threshold, i.e. for $Q \to 0$, $R_n$ has the expansion (e.g., \cite{bk}) 
\beq
R_n \simeq \frac{(2\pi^3)^{\frac{n-1}{2}}}{2\Gamma(\frac{3}{2}(n-1))}
\frac{(\prod_{j=1}^n m_j)^{1/2}}{(\sum_{j=1}^n m_j)^{3/2}} Q^{\frac{3n-5}{2}} 
\ . 
\label{rnthresh}
\eeq
Hence, near threshold, while the two-body phase space factor relevant for the
photofission reaction (\ref{photofission}) (before photon or prompt neutron
emission by the excited fission fragments) only involves a square root 
suppression $R_2 \propto Q^{1/2}$, the three-body 
phase space factor relevant for 
the pion production reactions (\ref{pi0}) and (\ref{pipm}) involves a more 
severe quadratic suppression, $R_3 \propto Q^2$.  A more detailed estimate 
would require calculation of the Coulomb distortion of the outgoing plane 
wave representing the pion, which would be represented by a factor analogous
to the Fermi function in nuclear beta decay.  However, the phase space
considerations discussed above suggest that it would be quite difficult to 
observe the pion production reactions near to their thresholds.  Slightly above
threshold, $\pi^0$ production by photofission 
on a heavy nucleus such as $^{238}$U
might show low energy $\pi^0$'s and higher energy $\pi^0$'s which have 
gained energy from the fission.  In addition, one should mention
the possibility that in the case where a $\pi^-$ is produced, it may form a 
Coulomb bound state with the larger-$Z$ fission fragment.  This channel would
suffer less kinematic suppression for two reasons: first, because it is a
two-body final state, and second because of the small additional release of 
energy due to
the Coulomb binding of the $\pi^-$.  However, the $\pi^-$ would rapidly be 
absorbed by the fission fragment to which it binds.

\section{``Sub-threshold'' Charged-Current Neutrino Reactions}

We next proceed to neutrino-induced fission reactions, namely the 
charged-current processes 
\beq
\nu_\mu + (Z,A) \to \mu^- + (Z+1,A) \to {\rm fission}
\label{nuz}
\eeq
and 
\beq
\bar\nu_\mu + (Z,A) \to \mu^+ + (Z-1,A) \to {\rm fission} \ . 
\label{nubarz}
\eeq
Suitable targets include $^{232}$Th, $^{238}$U, and the stable heavy nuclei
of Pb and Bi. 
The naive threshold energy for a charged-current (CC) reaction on a nucleus
$(Z,A)$ with incident $\nu_\mu$ (or $\bar\nu_\mu$) is 
\beq
E_{\nu,thr.} \simeq \Delta E_{Z \pm 1,Z;A} + m_\mu + \Delta E_{F,PB}
\label{eth}
\eeq
where $\Delta E_{F,PB}$ represents the effect of the Fermi momentum, $p_F
\sim 270$ MeV, of the struck nucleon and the effect of Pauli blocking. The
Fermi momentum smears out the threshold. We envision a situation in which, 
without fission, the incident energy
$E_\nu$ would be below threshold, but the reaction (\ref{nuz}) or 
(\ref{nubarz}) is rendered possible by energy transferred from the fission.
(Clearly, some of the 
energy transfer $q^0$ from the incident neutrino would be taken up 
to push the nucleus over the fission energy barrier of about 5 MeV.) 

In the case of an incident $\nu_\mu$, there would be a significant
probability for the resultant $\mu^-$ to form a Coulomb bound state
with one of the two fission fragments, preferentially the one with
higher $Z$. The $\mu^-$ could be captured in an
excited state and de-excite to the ground state with photon
emission.  The Coulombic binding energy, say in the ground state, is
$E_B \sim 5$ MeV for a fission fragment with $Z \sim 60$
\cite{fl}-\cite{eb}, so this would provide an additional source of
energy for the reaction.  If we assume that the mechanism of fission via the 
deformation of the nucleus \cite{bw} also applies to neutrino-induced fission
reactions, then the resultant time scale is expected to be similar to that
characterizing neutron-induced fission, of order $10^{-14}$ sec.  In
the fraction of the reactions where the muon does form a Coulomb bound
state with the nucleus of one of the fission fragments, the time
$\tau_\mu=\Gamma_\mu^{-1}$ during which the muon stays bound, before
it either decays or undergoes a CC reaction with the nucleus of the
fragment to which it is bound, is thus considerably longer than the
time characterizing the initial fission.

Near threshold, the usual charged-current neutrino cross sections on nuclei
vary strongly with energy.  For example, for the well-studied $A=12$ system,
the cross section for the reaction $^{12} C(\nu_\mu,\mu^-) ^{12}N$ 
has been calculated to vary from
less than $10^{-42}$ cm$^2$ slightly above the threshold of about 120 MeV to
$\sim 0.7 \times 10^{-40}$ cm$^2$ for $E_\nu \gsim 150$ MeV \cite{v12}.
Similar energy dependence is found in calculations of neutrino reactions on
heavier nuclei such as $^{56}$Fe and $^{208}$Pb \cite{kl00} (see also the
related works \cite{ls}-\cite{v3}) and would be expected to hold for the
analogous reactions on $^{232}$Th or $^{238}$U.  Since these reactions involve
a two-body final state, the phase space factor has the $Q^{1/2}$ dependence in
this threshold region as $Q \to 0$. In the 
present case, for the CC reactions yielding an outgoing $\mu^\pm$, the cross
section would be further suppressed near threshold, since the three-body phase
space factor by itself goes like $Q^2$ for $Q \to 0$.  However, for the
fraction of the reactions with an incident $\nu_\mu$ in which the $\mu^-$ forms
a Coulomb bound state with the nucleus of the higher-$Z$ fission fragment, one
would still have the less severe $Q^{1/2}$ factor.

  To obtain a rough estimate of the cross section for the reaction (\ref{nuz})
leading to a Coulomb bound state of the $\mu^-$ with the nucleus of a fission
fragment, with $E_\nu$ below threhold for the production of an asymptotic
$\mu^-$ state, but above the true threshold, taking into account the release of
fission energy and the smaller energy release from the Coulomb binding, we
proceed as follows.  The first step consists of the reaction $\nu_\mu + n \to
(p + \mu^-)_{virtual}$ on one of the neutrons in the nucleus. This elementary
process is characterized by a usual weak amplitude $\propto G_F$ and a time
$t_W \sim 1/m_W \sim 10^{-26}$ sec. Assuming that the energy transfer to the
nucleus is sufficient to push it over the fission barrier of about 5 MeV, 
the nucleus fissions
at a later time, around $t_{fiss.} \sim 10^{-14}$ sec.  The fission
energy can be transferred by the exchange of a virtual
photon.  To incorporate this in the amplitude one would multiply by a factor
$\sim Z \alpha$, and hence $\sim (Z \alpha)^2$ in the cross section.  Next, for
the formation of the Coulomb bound state, one would multiply the rate by a
factor $|\psi(0)|^2$ where $\psi$ is the quantum mechanical wavefunction
describing the $\mu^-$ and the nuclear fission fragment with charge $Z_f$ to
which it binds \cite{muattach}.  If this fission fragment were a point charge then, since the
Bohr radius $a$ of the ground state of the Coulomb bound state is much smaller
(by the factor $m_e/m_\mu=1/207$) than those of the electrons, there would not
be strong screening of the nuclear charge, and one would have simply
$|\psi(0)|^2 = (\pi a^3)^{-1} = (Z_f \alpha)^3 m_\mu^3/\pi$.  However, since
$a$ is comparable to the nuclear radius, one must take account of the
non-pointlike nature of the nuclear charge.  Combining the factor of $(Z_f
\alpha)^3$ with the factor of $(Z \alpha)^2$ for the energy transfer from the
fissioning nucleus, one has an overall factor of $f_{NIFM} \simeq
Z^2Z_f^3\alpha^5$, where $NIFM$ stands for neutrino-induced fission with
formation of a muonic Coulomb bound state. Substituting $Z=93$ for the virtual
$^{238}$Np nucleus resulting from the elementary reaction $\nu_\mu n \to \mu^-
p$, and a typical $Z_f=60$ for the larger-$Z$ fission fragment, this factor is
$f_{NIFM} \simeq 0.04$.  Therefore, a lowest-order estimate of the cross
section for the neutrino-induced fission reaction (\ref{nuz}) with $\mu^-$
binding to a fission fragment could be obtained by starting with the cross
section for the corresponding nonfission reaction, making the substitution
$E_\nu \to E_\nu + aE_{fiss.} + E_B$, and then multiplying by the factor 
$f_{NIFM}$, where for a given target nucleus $(Z,A)$ the value of 
$E_{fiss.}$ depends on which fission products are produced, and the factor 
$a$ represents the fraction of the fission energy available for transfer to 
the $\mu^-$.  Thus if one uses $^{238}$U as the target nucleus, and (i) 
$E_\nu$ is below the naive threshold by, say, 30 MeV, (ii) the fission
energy release is 150 MeV, of which about 80 MeV is transferred to the muon, 
and (iii) the muon Coulomb binding energy is 5 MeV, it follows that
the true energy is about 55 MeV above the true threshold, then, taking into
account the factor $f_{NIFM}$ with $Z_F \simeq 60$, our rough estimate of
the cross section for the neutrino-induced fission reaction leading to the
formation of a Coulomb bound state of the $\mu^-$ with the nucleus of the $Z
\sim 60$ fission fragment suggests that this cross section could be as large as
a few percent of the corresponding weak charged-current reaction a similar
interval of 55 MeV above threshold, and hence of order $10^{-43}- 10^{-42}$ 
cm$^2$.

One could also carry out a similar estimate for the reaction channels leading
to the production of an outgoing $\mu^\pm$.  For these channels, one would not
have to multiply by $|\psi(0)|^2$, but the cross section would be suppressed by
the more severe $Q^2$ dependence of the phase space near threshold for the
three-body final state.  Furthermore, if the effective energy is only slightly
above the true threshold, the $\mu^+$ from an incident $\bar\nu_\mu$ and the
$\mu^-$ from an incident $\nu_\mu$ (if the $\mu^-$ is not bound in a muonic
atom with a fission fragment) will have low energy, so that the Coulomb
correction to the rate will be substantial.  In the absence of fission, this
would be approximately given by the Fermi function 
\beq 
F(E,Z) = \frac{2\pi \eta}{1-e^{-2\pi \eta}}
\label{fez}
\eeq
where
\beq
\eta = \mp \frac{Z_f \alpha}{\beta}
\label{eta}
\eeq
where the $-$ ($+$) sign applies for an 
outgoing $\mu^+$ ($\mu^-$), $Z_f$ is again
the charge of the final state nucleus, and $\beta=v/c$ is the dimensionless
velocity of the outgoing $\mu^\pm$ \cite{schop,segre}.  This factor suppresses
(enhances) the emission of $\mu^+$ ($\mu^-$).  In the present case, the Coulomb
effect is more complicated to compute because the outgoing charged lepton
interacts not just with the 
Coulomb field of a single final-state nucleus, but with
the Coulomb fields of the two fission fragments, with two relative velocities,
$\beta_1$ and $\beta_2$.  However, the qualitative effect would be similar to
that for a charged lepton recoiling against a single nucleus.  

In estimating the cross sections for these ``below-threshold'' charged-current
neutrino reactions leading to an 
outgoing $\mu^\pm$, it is useful to observe that
in one respect they are analogous to a rare type of fission in which the final
state consists not just of the two daughter fission fragments (usually with
rather asymmetric mass distributions), but also a long-range $\alpha$ particle
or higher-$A$ cluster \cite{hyde,cluster}.  In these processes the energy of 
the fission products is partially transferred to other particles, e.g. 
long-range $\alpha$'s and ternary fission clusters.  The emission 
of energetic $\alpha$'s is peaked near to $90^\circ$ relative to the
axis of motion of the two outgoing fission fragments and hence is denoted as
``equatorial emission''.  A plausible explanation
of this process, within the general liquid-drop model of fission, is that as
the two lobes of the droplet are pulling apart, the $\alpha$ particle is
emitted from a region of the drawn-out neck between these two lobes, and the
$\alpha$ particle is then accelerated roughly away from the axis of the
receding droplets by the Coulomb repulsion with the nuclei of these 
fission fragments.  A typical
energy for the $\alpha$ particle is 20 - 30 MeV. For a heavy nucleus such as
$^{235}$U, this ternary fission with emission of an energetic $\alpha$ occurs
with a frequency, relative to the usual binary fission, of a few parts in
$10^3$.  This thus gives some measure of the suppression due to phase space and
Coulombic energy transfer.  A rarer type of ternary fission occurs when an
$\alpha$ particle or heavier cluster is emitted roughly along the polar axis 
defined by the two receding fission fragments, termed ``polar
emission'' \cite{cluster}.  This emission has been interpreted as being due in
part to the excitation of a giant dipole resonance \cite{cluster} due to the
fission process.  Estimates of the energy transfered
in the decay of this giant dipole resonance to the $\alpha$ particle or heavier
cluster are consistent with the observation that these latter particles tend to
have higher energies than the $\alpha$ particles emitted in an equatorial
manner.  

The phenomenon of ``below-threshold'' charged-current neutrino reactions on
heavy nuclei may also be relevant to supernova neutrinos. We recall that
neutrinos from supernovas would, in general, be comprised of a mix of
$\nu_\ell$ and $\bar\nu_\ell$, $\nu_\ell=\nu_e, \nu_\mu, \nu_\tau$. The
$\nu_\mu$ and $\bar\nu_\mu$ have average energies of about 25 MeV and flux
distributions, as a function of energy, that extend up to about 50 MeV
\cite{burrows}.  A method for observing these (anti)neutrinos via neutral
current reactions on $^{16}$O involving proton or neutron emission, populating
excited states of $^{15}$N and $^{15}$O which then undergo photon emission, has
been discussed in \cite{lvk}.  We note that although the above range of
energies of supernova (anti)neutrinos is below threshold for usual
charged-current $\mu^\pm$ production, this could be rendered possible by the
neutrino-induced fission that we have discussed here.  Thus, these
charged-current reactions could, in principle, help as a means for the
detection of supernova-generated $\nu_\mu$'s and $\bar\nu_\mu$'s, explicitly
distinguishing these from other types (flavors) of (anti)neutrinos, which
neutral current reactions do not do.  Although it is not straightforward to
calculate accurate cross sections for our neutrino-induced fission reactions
going beyond the rough estimates given here, and these cross sections may well
be somewhat smaller than those for the neutral current reactions of \cite{lvk},
our reactions have the useful property of yielding information on the type of
incident (anti)neutrino, $\nu_\mu$ or $\bar\nu_\mu$, and are subject to
different systematics. The detection of our reactions would use an appropriate
$^{238}$U target, from which the neutrons due to the fission (adding to those
from neutral current reactions) could be detected. Since the the time of the
initial arrival of neutrinos from the supernova would have been obtained from a
parallel neutrino detector containing hydrogen (e.g., in water) via the
$\bar\nu_e p \to e^+ n$ reaction, any background from spontaneous fission
during the $O(10)$ sec. duration of the supernova neutrino signal would be very
small.  For incident $\bar\nu_\mu$ one might also be able to detect the
outgoing $\mu^+$ via the high-energy positron from its decay.

\section{Concluding Remarks}

In conclusion, we have discussed several types of reactions that are naively
below threshold but can proceed because of the release of binding energy from
fission and the formation of Coulombic bound states.  These include
photofission reactions with pion production and charged-current neutrino
reactions on heavy nuclei.  It would be of interest to search for these
reactions experimentally.

\vspace{8mm}
 
The U.S. government retains a non-exclusive royalty-free license to
publish or reproduce the published form of this contribution or to
allow others to do so for U.S. government purposes.  This research was
supported in part by DOE DE-AC02-98CH10886.  The research of R.S. was
also supported by the NSF grant PHY-97-22101 at Stony Brook.

\vfill
\eject


\begin{thebibliography}{99}
\bibitem{eb}{Recall that if the nucleus of the fission fragment 
to which the $\mu^-$ binds
were a point charge, then the Coulombic binding energy in the lowest
Bohr orbit would be $E_B=(Z_{eff} \alpha)^2m_\mu/2$ where $Z_{eff}$
denotes the effective nuclear charge.  The lowest Bohr orbit
for this point charge case would be $a = 1/(Z_{eff}\alpha m_\mu)$, which
is comparable to the nuclear radius, so that there is not much 
shielding by electrons, but one must take account of the 
nonpointlike structure of the nuclear charge.  Extensive 
measurements of radiative de-excitations in muonic atoms have determined 
the muonic energy levels \cite{fl}-\cite{mukh}.}
\end{thebibliography}
\end{document}